\documentclass[sigconf,natbib=true]{acmart}

\AtBeginDocument{%
  }

\setcopyright{acmlicensed}
\copyrightyear{2018}
\acmYear{2018}
\acmDOI{XXXXXXX.XXXXXXX}
\acmConference[Conference acronym 'XX]{Make sure to enter the correct
  conference title from your rights confirmation email}{June 03--05,
  2018}{Woodstock, NY}
\acmISBN{978-1-4503-XXXX-X/2018/06}

\usepackage{amsmath}
\usepackage{graphicx}
\usepackage{booktabs}
\usepackage{multirow}
\usepackage{enumitem}

\begin{document}

\title{Deferred is Better: A Framework for Multi-Granularity Deferred Interaction of Heterogeneous Features}


\author{Yi Xu}
\authornote{Both authors contributed equally to this research and are co-first authors.}
\affiliation{
  \institution{Alibaba Group}
  \city{Beijing}\country{China}
}
\email{xy397404@alibaba-inc.com}
\author{Moyu Zhang}
\authornotemark[1]
\affiliation{
  \institution{Alibaba International Digital Commerce Group}
  \city{Beijing}\country{China}
}
\email{zhangmoyu.zmy@alibaba-inc.com}

\author{Chaofan Fan}
\affiliation{
  \institution{Alibaba Group}
  \city{Beijing}\country{China}
}
\email{fanchaofan.fcf@alibaba-inc.com}

\author{Jinxin Hu}
\affiliation{
  \institution{Alibaba Group}
  \city{Beijing}\country{China}
}
\email{jinxin.hjx@alibaba-inc.com}
\authornote{Corresponding author}

\author{Yu Zhang}
\affiliation{
  \institution{Alibaba Group}
  \city{Beijing}\country{China}
}
\email{daoji@alibaba-inc.com}

\author{Xiaoyi Zeng}
\affiliation{
  \institution{Alibaba Group}
  \city{Beijing}\country{China}
}
\email{yuanhan@taobao.com}

\renewcommand{\shortauthors}{Trovato et al.}

\begin{abstract}
Click-through rate (CTR) prediction models estimates the probability of a user-item click by modeling interactions across a vast feature space. A fundamental yet often overlooked challenge is the inherent heterogeneity of these features: their sparsity and information content vary dramatically. For instance, categorical features like item IDs are extremely sparse, whereas numerical features like item price are relatively dense. Prevailing CTR models have largely ignored this heterogeneity, employing a uniform feature interaction strategy that inputs all features into the interaction layers simultaneously. This approach is suboptimal, as the premature introduction of low-information features can inject significant noise and mask the signals from information-rich features, which leads to model collapse and hinders the learning of robust representations. To address the above challenge, we propose a \textbf{M}ulti-\textbf{G}ranularity Information-Aware \textbf{D}eferred \textbf{I}nteraction \textbf{N}etwork (MGDIN), which adaptively defers the introduction of features into the feature interaction process. MGDIN's core mechanism operates in two stages: First, it employs a multi-granularity feature grouping strategy to partition the raw features into distinct groups with more homogeneous information density in different granularities, thereby mitigating the effects of extreme individual feature sparsity and enabling the model to capture feature interactions from diverse perspectives. Second, a delayed interaction mechanism is implemented through a hierarchical masking strategy, which governs when and how each group participates by masking low-information groups in the early layers and progressively unmasking them as the network deepens. This deferred introduction allows the model to establish a robust understanding based on high-information features before gradually incorporating sparser information from other groups. Extensive offline and online experiments validate the superiority of MGDIN.
 
\end{abstract}


\maketitle
\section{Introduction}
Click-through rate (CTR) prediction is a cornerstone task in recommendation systems and a critical determinant of user experience\cite{Zhang_2019,intro2,DIN}. Its core challenge lies in effectively modeling the complex and subtle interactions across a vast and diverse feature space to accurately estimate the probability of a user clicking on a candidate item\cite{liu2024collaborativeensembleframeworkctr,wukong,DHEN}. The advent of deep learning has spurred the development of feature interaction architectures in the CTR field, such as Factorization Machines\cite{FM,DeepFM,AFM,FFM}, Cross-Networks\cite{dcn,Wang_2021,DCNv3}, and Attention mechanisms\cite{vaswani2023attentionneed,AutoInt,MTGR}. These models aim to automatically learn high-order feature combinations, moving beyond manual feature engineering to generate expressive and high-quality hidden representations that enhance models' predictive performance\cite{EulerNet}.

However, despite the remarkable achievements of existing deep CTR prediction models, they are almost universally built upon a uniform feature interaction strategy, i.e., "one-size-fits-all". This approach inputs all features into the feature interaction layers simultaneously, ignoring a fundamental yet often overlooked challenge: feature heterogeneity. Unlike in fields such as computer vision (CV) or natural language processing (NLP), the features in recommender systems are inherently heterogeneous, exhibiting significant disparities in statistical properties such as sparsity, cardinality, and information content\cite{hstu,interformer,collapse,gui2023hiformerheterogeneousfeatureinteractions,zhai2022scalingvisiontransformers,LLaMA}. For example, the extreme sparsity of an Item ID feature is fundamentally different from the dense nature of a numerical feature like user age and item price, or a low-cardinality categorical feature like device type. 

While numerous recent studies seek to mitigate the issue of item ID sparsity by augmenting them with multimodal semantics, they frequently neglect the heterogeneous information density across other features, which in turn creates performance bottlenecks. The premature introduction of noisy and undertrained feature representations from extremely sparse features can inject significant noise into the feature interaction process. This interaction noise often overwhelms the robust signals from more informative and dense features, leading to suboptimal representations or even catastrophic representation collapse, ultimately hindering model convergence and degrading final prediction performance.

To overcome the above limitations, we challenge the fundamental premise of synchronous feature interaction and propose a new guiding principle: deferred is better. Our core insight is that features should not participate uniformly at all stages of the interaction process. Instead, we argue that a more robust model can be built by first establishing a stable foundation using high-information features, and only then gradually incorporating sparser, more granular, and potentially noisier features. Acting on this principle, this paper introduces a \textbf{M}ulti-\textbf{G}ranularity Information-Aware \textbf{D}eferred \textbf{I}nteraction \textbf{N}etwork (MGDIN), which adaptively defers the incorporation of each feature group into the interaction process. 

Specifically, MGDIN's core mechanism operates through two distinct phases: 1) Multi-Granularity Feature Grouping. To address the uneven information content of features, we first employ a multi-granularity grouping strategy, which partitions the raw features into distinct groups with more homogeneous information density. The goal is twofold: to maximize the collective information content within each group and to mitigate the disruptive impact of extremely sparse individual features on subsequent model interactions. 2) Information-Aware Deferred Interaction. Building on the above feature groups, we then implement a deferred interaction mechanism via a hierarchical masking strategy. This strategy governs the timing of each group's participation based on its information content. In the early layers, low-information groups are masked to prevent representation collapse arising from premature and noisy interactions. As the network deepens, these masks are progressively removed, allowing the groups to participate in the feature interaction process. This deferred introduction operation enables the model to first establish a robust understanding from high-information features before gradually and safely integrating the sparser information from other feature groups.

Our contributions are summarized as follows:
\begin{itemize}
    \item To the best of our knowledge, we are the first work to propose a deferred feature interaction framework to address the challenge of feature heterogeneity in CTR prediction.
    \item By employing a multi-granularity strategy and deferred interaction, MGDIN mitigates the issues arising from feature heterogeneity, namely, uneven information distribution and representation collapse due to sparse features, enabling the model to learn a more robust feature interaction paradigm.
    \item Comprehensive evaluations, encompassing both offline datasets and online A/B tests, not only confirm the superiority of MGDIN over state-of-the-art baselines but also underscore the efficacy of its core deferred interaction principle.
\end{itemize}

\section{Preliminaries} 
Given a user $u$ and a candidate item $i$ with a set of features $\mathcal{X}$ under a specific context, the input features encompass user profiles (e.g., age, gender), item attributes (e.g., item ID, category ID, brand ID), and contextual information (e.g., geographic location, device). To address the extreme sparsity of item IDs, we leverage recent advancements by replacing the original item IDs with dense semantic IDs generated from both multimodal content and user behavior. This foundational step substantially mitigates heterogeneity at the input level. The input of the model can be represented as $\mathcal{X}=[x_1,\dots,x_n]$, where $n$ is the number of features.

The objective of CTR prediction is to estimate the probability $\hat{y} \in [0, 1]$ that the user $u$ will click the item $i$ under $\mathcal{X}$, which can be formulated as follows, where $\theta$ denotes the model parameters:
\begin{equation}
\label{eq:task}
    \hat{y} = f(\mathcal{X}; \theta)
\end{equation}

\section{Methodology} 


While prior work has mitigated the most extreme form of heterogeneity by introducing semantic IDs for item IDs, this only partially solves the feature heterogeneity problem. Significant information imbalances persist among the remaining features. Therefore, continuing to employ a "one-size-fits-all" interaction paradigm is suboptimal, which will make the model remain vulnerable to noise from sparser features, trigger premature representation collapse, and prevent the learning of effective feature interaction.

To address the above challenge, we propose the Multi-Granularity Information-Aware Deferred
Interaction Network (MGDIN), illustrated in Figure 1. MGDIN's architecture is composed of two core components:1)Multi-Granularity Feature Grouping, which applys a grouping strategy that partitions the features into distinct groups with more homogeneous information density to minimize the disruptive influence of extremely sparse features on subsequent interactions; 2)Information-Aware Deferred Interaction, which implements a delayed interaction mechanism to prevent premature noisy interactions from causing representation collapse by adaptively controlling when each feature group participates in the interaction.

\begin{figure}[t]
\includegraphics[width=0.9\columnwidth]{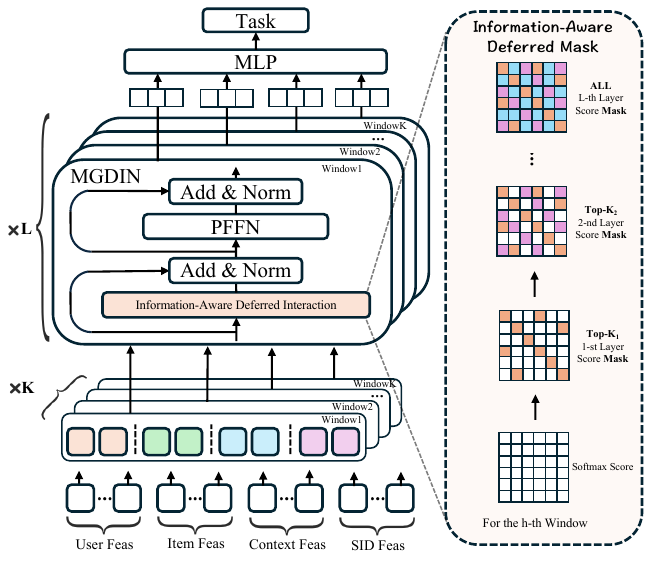}
\centering
\caption{The network architecture of the Multi-Granularity Information-Aware Deferred
Interaction Network (MGDIN).}
\label{fig:intro_heat}
\vspace{-0.3cm}
\end{figure}


\subsection{Multi-Granularity Feature Grouping}
To address the uneven distribution of information content across features, we divide the original features into groups with a more uniform information density using feature windows. However, recognizing that different window partitions can help capture feature interactions at various granularities and from multiple perspectives, we adopt a multi-granularity feature grouping approach in this paper. Specifically, we partition the original features into distinct sets of groups using multiple window sizes, and the feature groups corresponding to a single window size interact with one another in the same window size. The computations for the different window granularities are performed in parallel, as shown in the lower half of Figure 1. Each window, in parallel, processes the feature groups of its corresponding granularity and their subsequent interactions. Since the computations involving the feature groups within each window are similar, for simplicity, we will only illustrate the process for a single window granularity in the calculations that follow.

For the $h$-th window branch, we denote the grouping granularity by $g_h$, and the original feature set $\mathcal{X}$ is partitioned into a set of feature groups based on this granularity:
\begin{equation}
     {\mathcal{X}^{h}= [f_1, f_2, .., f_{\frac{n}{g_h}}]}
\end{equation}
where $f_1 = x_1 \oplus x_2 \oplus \cdots \oplus x_{g_h}$. Following this process for $K$ windows, we thus obtain $K$ distinct sets of feature groups, $\left\{ \mathcal{X}^{1}, ..., \mathcal{X}^{h}, ..., \mathcal{X}^{K} \right\}$, where feature interactions will be computed within each set in subsequent modules. Thus, with the feature groups having a relatively uniform information content, feature interactions can be captured from diverse perspectives through this multi-granular approach.

\subsection{Information-Aware Deferred Interaction}
As mentioned above, to more effectively alleviate the representation collapse problem stemming from feature heterogeneity during feature interaction, we propose to use a hierarchical masking strategy to delay the interaction of sparse features, which gradually incorporates information from sparser feature groups, thereby promoting robust feature interaction. Recognizing that feature representations gradually incorporate information from other features as the network depth increases, which can cause attention scores in deeper layers to lose their guiding significance regarding information content, we calculate the attention score matrix using the initial representations of the feature groups from the model's first layer. Then, based on these scores, we progressively introduce feature interactions in ascending order of their corresponding scores, thereby ensuring that features with lower information content do not disrupt feature interactions in the early stages. 

Specifically, for the $h$-th window granularity, we calculate the attention scores between feature groups at this granularity and obtain the attention score matrix, as shown below:
\begin{equation}
\mathbf{A}^{h}_0=(\mathcal{X}^{h}_0\mathbf{Q}^h_0) (\mathcal{X}^{h}_0\mathbf{K}^{h}_0)^{\top}
\end{equation}
where $\mathbf{A}^{h}_0 \in \mathbb{R}^{\frac{n}{g_h} \times \frac{n}{g_h}}$ represents the score between each pair of feature groups under the $h$-th window granularity. $\mathcal{X}^{h}_0$ denotes the original features grouped by $h$-th window granularity. Since sparse features have less information content and participate in fewer training times, the attention score here is often also very low. $\mathbf{Q}^h_0$ and $\mathbf{K}^{h}_0$ are the learnable parameters. Based on the learned attention score matrix, we can activate only the high-scoring feature interactions at each layer, as shown in the following calculation:
\begin{gather}
\mathbf{Z}_l^{h} = \left[ \left( (\mathcal{X}^{h}_{l-1}\mathbf{Q}^h_l) (\mathcal{X}^{h}_{l-1}\mathbf{K}^{h}_l)^\top \right) \odot \mathbb{I} \left( \mathbf{A}^{h}_0 \in \text{Top-}k_l \right) \right] \mathcal{X}^{h}_{l-1}\mathbf{V}^h_l \\
k_l = \left\lfloor \frac{l}{L} \cdot n^2 \right\rfloor, \quad l \in \{1, \dots, L\}
\end{gather}
where $\mathbf{Z}_l^{h}$ denotes the ouput feature sets of the $l$-th interaction layer under $h$-th window size. $\left( (\mathcal{X}^{h}_{l-1}\mathbf{Q}^h_l) (\mathcal{X}^{h}_{l-1}\mathbf{K}^{h}_l)^\top \right)$ denotes the attention scores of feature pairs in the $l$-th layer. $\mathbb{I}(\cdot)$ is the indicator function that outputs $1$ if the condition is satisfied and $0$ otherwise. $\mathbf{Q}^h_0$, $\mathbf{K}^{h}_0$, and $\mathbf{V}^{h}_0$ are the learnable parameters.
$k_l$ is the number of activations in the attention score matrix of each layer 
$l$, which gradually approaches full activation as the layer index $l$ increases

Subsequently, the residual connection operation and the Per-token Feed-Forward Networks (PFFN)\cite{onetrans} are employed to $\mathbf{X}_l^h$ to facilitate feature interactions further, as follows:
\begin{align}
& \hat{\mathbf{Z}}_{l}^h=\text{LN}(\mathbf{Z}_l^h+{\mathcal{X}}_{l-1}^h)\\
\mathcal{X}_l^h &= \text{LN}(\text{PFFN}(\hat{\mathbf{Z}}_{l}^h) + \hat{\mathbf{Z}}_{l}^h)
\end{align}
where $\text{LN}(\cdot)$ denotes the Layer Normalization operation.

\subsection{Prediction and Model Learning}
Following the operations described above, we can compute $K$ final outputs, denoted as $\left\{\mathcal{X}_L^1, ..., \mathcal{X}_L^h, ..., \mathcal{X}_L^K \right\}$, in parallel. After applying pooling operations to these $K$ sets of feature groups, we feed them into a prediction layer, thereby capturing information from diverse perspectives through feature interactions at multiple granularities:
\begin{equation}
\hat{y} = MLP(Pooling(\mathcal{X}_L^1, ..., \mathcal{X}_L^h, ..., \mathcal{X}_L^K))
\end{equation}
The optimization objective of MGDIN is the cross-entropy loss:
\begin{equation}
    \mathcal{L} = -\frac{1}{|\mathcal{D}|} \sum_{u,i \in \mathcal{D} }^{|\mathcal{D}|} \left( y \log(\hat{y}) + (1 - y) \log(1 - \hat{y}) \right)
\end{equation}
where $\mathcal{D}$ represents the training dataset, and $y \in \{0, 1\}$ indicates the ground truth label for the corresponding interaction.
\section{Experiments}
In this section, we conduct extensive experiments on public and industrial datasets to evaluate the effectiveness of MGDIN.
\subsection{Experimental Setup}
\subsubsection{Dataset}
To validate the effectiveness of MGDIN, we conduct experiments on two real-world large-scale datasets. 

\textbf{Amazon Sports}. It is derived from the Amazon platform. Specifically, we utilize the Sports subset and follow the preprocessing protocol established in \cite{amazon}, where we treat ratings greater than 3 as positive labels and those less than or equal to 3 as negative labels in Amazon Product Reviews. The first 90\% of the data are used for training, while the remaining 10\% constitute the test set.

\textbf{Industrial Dataset}: This dataset contains 7 billion user interaction records from an international e-commerce advertising system, featuring diverse item features and user behavior sequences.
\subsubsection{Evaluation Metrics} For the evaluation, we use the widely used AUC, GAUC, and LogLoss for prediction accuracy.

\sloppy
\subsubsection{Baselines} To validate the effectiveness of MGDIN, we evaluate it with state-of-the-art models, including FM\cite{FM}, DNN, Wide\&Deep\cite{widedeep}, DeepFM\cite{DeepFM}, DCN\cite{dcn}, AutoInt\cite{AutoInt}, GDCN\cite{gdcn}, MaskNet\cite{masknet}, PEPNet\cite{PEPNet}, RankMixer\cite{rankmixer}, and OneTrans\cite{onetrans}.

\subsubsection{Implementation Details} For the industrial dataset, MGDIN is configured with 3 layers and 4 windows, with granularities set to $\{32, 64, 96, 128\}$ and layer-wise sparsity ratios of $\{0.33, 0.66, 1.0\}$. For the Amazon-Sports dataset, we use 3 layers and 2 windows with granularities $\{2, 4\}$, maintaining the same sparsity levels.

\begin{table}[htbp] %
  \centering
  \caption{Overall performance comparison on two datasets. "Improv." denotes the relative improvement of MGDIN over the best baseline. The best baseline is denoted in \underline{underline}.}
  \label{tab:performance_overall}
  \resizebox{\columnwidth}{!}{%
    \begin{tabular}{llcccccc} 
      \toprule
      \multicolumn{2}{c}{\textbf{Dataset}} & \multicolumn{3}{c}{\textbf{Amazon Sports}} & \multicolumn{3}{c}{\textbf{Industrial}} \\
      \cmidrule(lr){3-5} \cmidrule(lr){6-8}
      \multicolumn{2}{c}{\textbf{Method}} & \textbf{AUC} & \textbf{GAUC} & \textbf{Logloss} & \textbf{AUC} & \textbf{GAUC} & \textbf{Logloss} \\
      \midrule
       \multicolumn{2}{l}{FM} & 0.6659 & 0.6427 & 0.3610 &0.6857&0.5893&0.1125 \\
       \multicolumn{2}{l}{DNN}  & 0.6695 & 0.6445 & 0.3594 & 0.6888&0.5901&0.1122  \\
       \multicolumn{2}{l}{Wide\&Deep} & 0.6678 & 0.6449 & 0.3581 &0.6875&0.5888&0.1124  \\
       \multicolumn{2}{l}{DeepFM} & 0.6747 & 0.6503 & 0.3541 & 0.6848&0.5860 &		0.1125  \\
       \multicolumn{2}{l}{DCN} & 0.6760 & 0.6507 & 0.3542 & 0.6915&0.5914&0.1121  \\
       \multicolumn{2}{l}{AutoInt} & 0.6730 & 0.6484 & 0.3699 &0.6874&0.5877 & 0.1124 \\
       \multicolumn{2}{l}{GDCN} & 0.6738 & 0.6495 & 0.3558 & 0.6894&0.5899&0.1123  \\
    \multicolumn{2}{l}{MaskNet} &0.6732 & 0.6446 & 0.3569  &0.6881&0.5876&0.1123  \\
      \multicolumn{2}{l}{RankMixer} & 0.6736 & 0.6501 & 0.3541 & \underline{0.6956}& 0.5916& \underline{0.1118}  \\
      \multicolumn{2}{l}{OneTrans} & \underline{0.6775} & \underline{0.6517} & \underline{0.3540} &\underline{ 0.6956}& \underline{0.5921}& \underline{0.1118}  \\
      \midrule			
       \multicolumn{2}{l}{\textbf{MGDIN}} & \textbf{0.6789} & \textbf{0.6544} & \textbf{0.3539} & \textbf{0.6994}&\textbf{0.5936}&\textbf{0.1116}\\ 
       \multicolumn{2}{l}{\textbf{Improv.}} &\textit{ +0.21\% }& \textit{+0.41\%} &\textit{-0.01\%}& \textit{+0.54\%} &\textit{+0.25\%} & \textit{-0.02\%} \\
      \bottomrule
    \end{tabular}%
  }
\end{table}

\subsection{Overall Performance} Table \ref{tab:performance_overall} presents the overall predictive performance of all methods on both industrial and public datasets, along with the relative improvement of our method over the best-performing baselines. The experimental results on both datasets robustly demonstrate the effectiveness of our method, MGDIN. Specifically, methods like Rankmixer and OneTrans can alleviate feature heterogeneity to some extent by aggregating feature groups, and they outperform traditional feature interaction models like AutoInt in terms of predictive performance. However, while they mitigate feature heterogeneity at a foundational level, they fail to effectively address the representation collapse problem caused by prematurely incorporating low-information features into the interaction process. Furthermore, they are limited to a single, fixed grouping strategy for feature interactions, thus failing to effectively model these interactions from multiple granularities and perspectives. In contrast, our proposed method outperforms these baselines, which underscores the effectiveness of our multi-granularity, information-aware delayed interaction mechanism and validates its underlying rationale.

\subsection{Ablation Study}
We conducted an ablation study with two variants: w/o MG, which uses a single window for feature grouping, and w/o DI, which replaces the delayed interaction mechanism with standard attention while retaining multi-granularity grouping. The experimental results are presented in Table \ref{tab:tiger_ablation}. As observed, MGDIN significantly outperforms both variants, which confirms the effectiveness of each proposed component. Specifically, removing w/o MG led to a notable degradation in both AUC and GAUC, highlighting the necessity of multi-granularity windows for capturing feature interactions from diverse perspectives. w/o DI also resulted in a significant performance drop, thus validating our hypothesis that even though MG helps establish a more uniform information distribution, residual heterogeneity persists. This suggests that having all features participate in every interaction stage is suboptimal, and for sparse feature groups, a deferred strategy is more effective.

\begin{table}[htbp]
  \centering
  \caption{Ablation study of MGDIN.}
  \vspace{-0.3cm}
  \label{tab:tiger_ablation}
  \resizebox{0.6\columnwidth}{!}{%
  \begin{tabular}{lccc}
    \toprule
    \multicolumn{1}{l}{\textbf{Variants}} & \textbf{AUC} & \textbf{GAUC} & \textbf{Logloss} \\
    \midrule
     \textbf{MGDIN} & 0.6994&0.5936&0.1116\\
       w/o MG & 0.6964 & 0.5910 & 0.1118 \\ 		 
       w/o DI & 0.6971 & 	0.5917& 0.1118\\ 	 		
    \bottomrule
  \end{tabular}
  }
\end{table}
\subsection{Hyper-parameters Analysis}
In this section, we study the impact of key hyperparameters on model performance. (a) Impact of the number of granularities $K$: As shown in Figure 2(a), model performance generally improves as the number of granular windows increases. This highlights the effectiveness of the multi-granularity approach in capturing feature interactions from diverse perspectives and modeling their complex patterns. (b) Impact of the deferred ratio $k_l$: While maintaining a progressive order from dense to sparse features, we varied the proportion of features incorporated at each interaction layer. As shown in Figure 2(b), the results indicate that these variations have a minimal impact on model performance, which aligns with our expectations. Since relatively sparse features do not constitute the majority within the dataset, the key is to ensure that their introduction is deferred to the later stages and prevent their noise from disrupting the crucial feature interactions in the early layers. 

\begin{figure}[t]
\includegraphics[width=0.8\columnwidth]{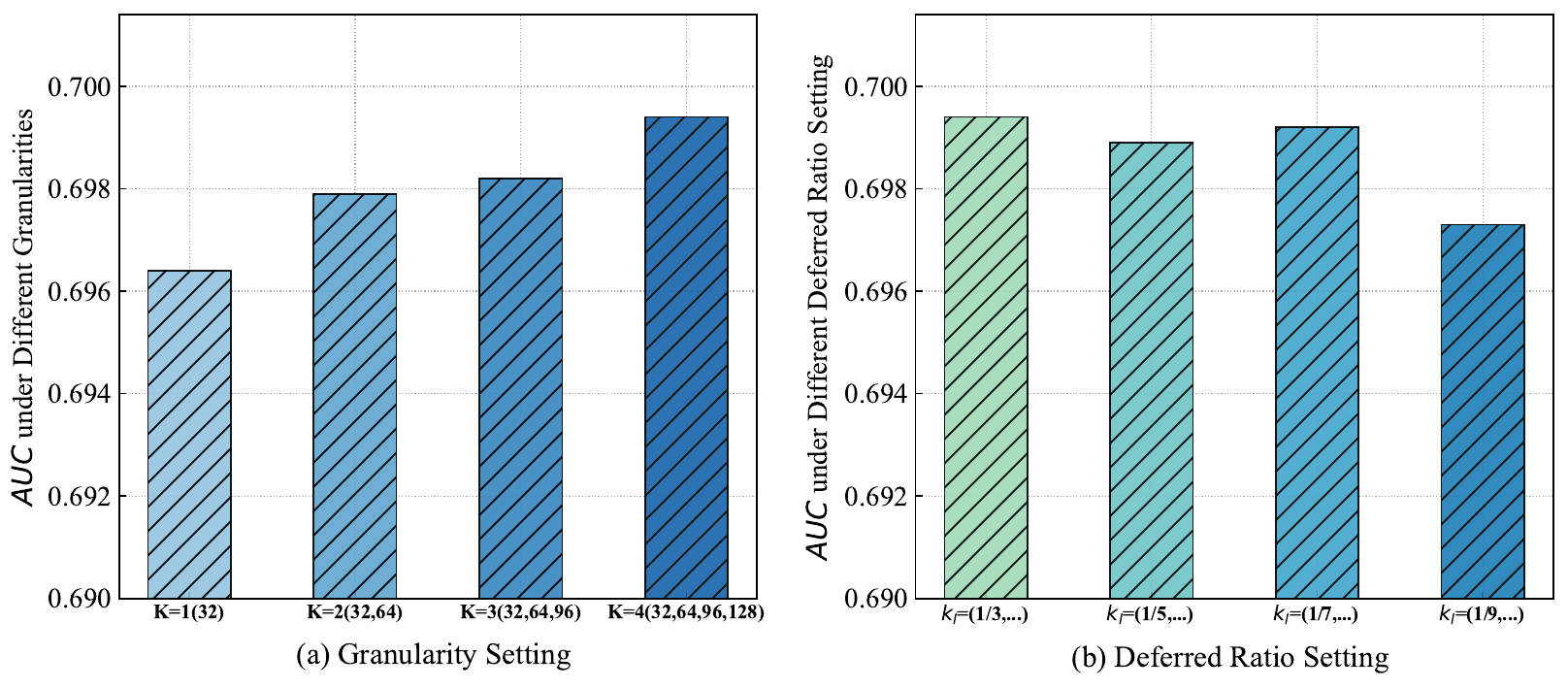}
\centering
\caption{The hyper-parameters analysis of MGDIN.}
\label{fig:intro_heat}
\vspace{-0.3cm}
\end{figure}

\section{Online Experiments}
To evaluate its real-world effectiveness, MGDIN was deployed in a 10-day online A/B test on a major e-commerce platform. Compared to the incumbent production model, MGDIN achieved a 3.04\% relative lift in CTR while incurring no additional inference latency. 

\textbf{Complexity Efficiency}: For our \textit{MGDIN}, the space complexity is $\sum_{h=1}^K(n/g_h)^2$, which is typically less than $O(n^2)$ of standard attention-based methods. Meanwhile, the time complexity is $max\left\{(n/g_h)^2\right\}_{h=1}^K$ due to the parallel operation.

\section{Conclusion}
This paper proposes the Multi-Granularity Information-Aware Delayed Interaction Network (MGDIN), designed to mitigate the feature interaction collapse problem stemming from feature heterogeneity. Specifically, MGDIN first partitions features into multiple groups at different granularities, which balances the information distribution and mitigates the disruptive impact of extremely sparse features. Subsequently, MGDIN determines the optimal timing for incorporating each feature group into the feature interaction process, which prevents feature interaction representation collapse by deferring the introduction of sparser features and allows the model to establish a robust understanding from high-information features before gradually integrating the sparser information from other feature groups. Finally, extensive experimental results demonstrate that the deferred strategy in MGDIN is more effective.
\bibliographystyle{ACM-Reference-Format}
\bibliography{main}
\end{document}